# Conditional Neural Process for non-parametric modeling of AGN light curves

Iva Čvorović - Hajdinjak*[1] | Andjelka B. Kovačević[1,2] | Dragana Ilić[1,3] | Luka Č. Popović[1,2,4] | Xinyu Dai[5] | Isidora Jankov[1] | Viktor Radović[1] | Paula Sánchez-Sáez[6] | Robert Nikutta[7]

[1]Department of Astronomy, Faculty of Mathematics, University of Belgrade, Studentski trg 16, 11000 Belgrade, Serbia
[2]Fellow of Chinese Academy of Sciences President's International Fellowship Initiative (PIFI) for visiting scientist
[3]Humboldt Research Fellow, Hamburger Sternwarte, Universitat Hamburg, Gojenbergsweg 112, 21029 Hamburg, Germany
[4]Astronomical Observatory Volgina 7, 11000 Belgrade, Serbia
[5]Homer L. Dodge Department of Physics and Astronomy, University of Oklahoma, Norman, OK 73019, USA
[6]Millennium Institute of Astrophysics (MAS), Nuncio Monse nor Sotero Sanz 100, Providencia, Santiago, Chile
[7]NSF's NOIRLab, 950 N. Cherry Ave, Tucson, AZ 85719, USA

**Correspondence**
* Email: iva.cvorovic@gmail.com

The consequences of complex disturbed environments in the vicinity of a supermassive black hole are not well represented by standard statistical models of optical variability in active galactic nuclei (AGN). Thus, developing new methodologies for investigating and modeling AGN light curves is crucial. Conditional Neural Processes (CNPs) are nonlinear function models that forecast stochastic time-series based on a finite amount of known data without the use of any additional parameters or prior knowledge (kernels). We provide a CNP algorithm that is specifically designed for simulating AGN light curves. It was trained using data from the All-Sky Automated Survey for Supernovae, which included 153 AGN. We present CNP modeling performance for a subsample of five AGNs with distinctive difficult-to-model properties. The performance of CNP in predicting temporal flux fluctuation was assessed using a minimizing loss function, and the results demonstrated the algorithm's usefulness. Our preliminary parallelization experiments show that CNP can efficiently handle large amounts of data. These results imply that CNP can be more effective than standard tools in modeling large volumes of AGN data (as anticipated from time-domain surveys such as the Vera C. Rubin Observatory's Legacy Survey of Space and Time).

## 1 | INTRODUCTION

Active Galactic Nuclei (AGN) are among the most powerful and compact sources in the Universe. On time spans ranging from years to hours, significant luminosity variations are typical. Because of the high luminosity and short variable time scale, AGN power is generated via events that are more efficient in terms of energy release per unit mass than regular star processes (Fabian, 1997). The presence of supemassive black hole (SMBH) in the core of AGN has been theoretically inferred based on the fact that high radiative efficiencies can be obtained by accreting matter onto a black hole or extracting its spinning energy (Rees, 1984; Rees, Begelman, Blandford, & Phinney, 1982). AGN model comprises an accretion disk orbiting an SMBH, clouds of both broad and narrow-line emitting gas (i.e., Broad Line Region - BLR and Narrow Line Region - NLR), and a dusty region that obscures the disk component when viewed close to edge on (Elvis, 2000; Urry & Padovani, 1995). The accretion disk's continuum radiation photoionizes gas clouds around it (i.e. the BLR), resulting in broad emission lines that encode the clouds' geometry and kinematics (Krolik, 1999; Osterbrock, 1989; Peterson, 1997). The physical approach to reverberation mapping is simple (see e.g., Peterson, 2006). The need to learn about the physical mechanisms powering AGN over a wide range of observable timescales has sparked interest in improving the statistical models to extract information encoded in a stochastic light curve. Long-term monitoring campaigns produced an important result that the BLR spans light hours (Cho et al., 2021) to hundreds of light days, depending on luminosity (e.g., Bentz et al., 2013; Denney, Peterson, & Pogge, 2009; Peterson, 1998; Shapovalova et al., 2019; Wang, Songsheng, Li, Du, & Zhang, 2020).

Thus, optical variability has long been a diagnostic of the processes in the central engines of active galactic nuclei (Antonucci, 1993, Bonfield et al., 2010, Richards et al., 2006, Shang et al., 2011). The exploration of the variability of AGN light curves has led to understanding underlying physical process in AGNs (Ulrich et al., 1997, Wagner et al., 1995).

Variability might be categorized into three classes based on their changeable characteristics: periodic, stochastic, and explosive. Periodic phenomena are caused by orbital motion, rotation, precession and pulsation (e.g. binary/multiple black holes, jets). Accretion onto black holes, and rare processes (e.g., scintilation) are all examples of stochastic events. Thermonuclear (X-ray bursts); magnetic reconnection (flares), and tidal disruption events are examples of explosive phenomena. The main applications of light curve analysis include detection of periodic motion, variable objects analysis, and classification.

Since optical AGN variability studies are data driven rather than theory driven, statistical modeling is a practical tool that, at minimum, can be used to also to constrain theoretical models.

One important problem so frequently encountered in almost all varaible objects is that the sampling is on the same order as harmonic of period and aliasing can appear. Randomizing the sampling interval can sometimes break these aliases and help in period detection. Another serious problem is that there are often large seasonal gaps, due to the annual visibility of the target (Shore, Livio, & van den Heuvel, 1994). These often contribute uncertainty to periods and there are no simple solutions to these problems (Shore et al., 1994).

For example, periodicity could indicate the presence of a binary system (Graham et al., 2015; Kovačević et al., 2019) however it requires high sampling frequency, since the true characteristic frequencies could be higher than the sampling frequency and thus the reconstructed model would be distorted to some extant (Moreno et al., 2019).

Furthermore, with realistic light curve model, we can distinguish AGNs that experience changing look events (Oknyansky et al., 2018; Shapovalova et al., 2019; Ilić et al., 2020; Sánchez-Sáez et al., 2021), as well as the cutoff in the X-ray fluctuation in power density spectrum (Edelson et al., 1999), monochromatic flares (Cid Fernandes, Sodre, & Vieira da Silva, 2000), possible thermal fluctuations driving optical ones the fluctuations in light curves of accreting black holes (Kelly, Bechtold, & Siemiginowska, 2009), or even possibility that both the optical and X-ray fluctuations are caused by inwardly propagating fluctuations in, say, the accretion rate (Kelly, Sobolewska, & Siemiginowska, 2011)

Besides typically being sparse, gapped and irregular, AGN light curves are also nonlinear and stochastic. All of these properties make them difficult to model and analyze. To determine underlying characteristics of a light curve source, such as estimating the time lag from the cross-correlation analysis or detecting possible periodic or quasi-periodic behaviour, we have attempted to model available observed light curves as a means to pre-process and prepare them for further analysis. Modeling of AGN light curves eliminates the problem of erratic cadence, gaps, and other issues, due to reliably generated data for any point in time covered by the modeled light curve.

There are several approaches for modeling of AGN light curves, such as Damped Random Walk (DRW, Kelly et al., 2009; Kozłowski, 2017; Sánchez-Sáez et al. 2018), Gaussian Process (GP) with different kernels (Kovačević et al., 2019; Shapovalova et al., 2017; Shapovalova et al., 2019) or Autoencoders (Tachibana et al 2020). Initial application of Gaussian processes (Kovačević et al., 2021) shows that machine learning can be useful for time lag and periodicity detection for light curves with cadences above 100 observations per 10 years. The optical variability of AGNs has been most commonly represented by a stochastic model based on DRW process (Kelly et al., 2009). The DRW model incorporates two hyper-parameters: a characteristic amplitude $\sigma$, which affects exponentially decaying variability with time scale $\tau$ around the mean magnitude $m_0$ of AGN light curve. These two important aspects of DRW affect AGN light curve modeling (Kozłowski, 2021).

The Conditional Neural Process (CNP, Garnelo et al., 2018) for modeling AGN light curves, which we here introduce, does not require any additional parameters or prior assumptions on mathematical structure of light curve data, unlike GPs. The reason we examined the applicability of this particular method is the stochastic nature of AGN data and the fact that CNP incorporates ideas from Gaussian Process, which are already successfully used for AGN light curve modeling, into a neural network training regime (Garnelo et al., 2018). Furthermore, GPs have scaling issues, especially with sparse data, which require additional computational resources for calculating the most suitable kernel function or switching between more kernels for more complex problems. These types of regression tasks are not trivial for GPs, whereas for CNP this could be solved merely by changing the data set used for training (Garnelo et al., 2018). Therefore, the complexity of modeling process is reduces from $O((n + m)^3)$, where $n$ is the number of known data points and $m$ is the number of unlabeled data points, characteristic to GPs, to $O(n + m)$ which is achieved with CNP (further discussed in Section 5). Hence, deep learning could be highly desirable in the era of big AGN data (Tachibana et al., 2020). The next generation time domain surveys, such as Vera Rubin Observatory Legacy Survey of Space and Time (LSST, Abell et al., 2009, Ivezić et al., 2019), will provide observations with different cadences over ten years for millions of AGN sources (Brandt et al., 2018; Bianco et al., 2021) in six filters - *ugrizy*. This will result in an immense amount of data, which will require efficient tools for processing. Therefore, aside from determining applicability of CNP for modeling light curves, we have also made initial parallelization tests of our code on a small scale with High-Performance Computing (HPC) system, to ascertain the next steps in optimizing performance of our code, so that it can be used in processing large amount of data, expected in the future time domain surveys.

To determine the applicability of Conditional Neural Process we took a sample of 153 AGNs from the All-Sky Automated Survey for Supernovae (ASAS-SN, Kochanek et al., 2017; Shappee et al., 2014), whose characteristics are described in Section 2. Our main task has been implementation and optimization of CNP. The architecture of CNP is described in Section 3, along with AGN light curve properties, used in CNP modeling. In Section 4 we present detailed results from testing five representative objects from our research. We discuss initial results for modeling and optimization, and define the next steps for future research in Section 5, concluding that CNP could be used for modeling AGN light curves and indicating possible scope for improvement, using parallelization and other methods for code optimization.

## 2 | DATA

The CNP algorithm has been trained and tested on a sample of 153 AGNs (see Figure 1), that were detected in the first nine months of the all sky survey by the Swift Burst Alert Telescope (9-Month BAT Survey[1], Tueller et al., 2008). Their optical light curves were taken from the ASAS-SN database (Holoien et al., 2017), and they cover more than 2000 days.

We have selected this set of AGN light curves for several reasons. They have a good representation of 80% sky coverage, and show particularities, such as flares, possible quasi-periodic oscillations, time gaps, and irregular cadence, enabling us to test our tool with challenging data sets.

Our sample of AGN light curves is thus inhomogenous in many aspects. For example, observational coverage of each source is between 100 and 600 data points (Figure 2). The distribution is bi-modal with large right tail, indicating unfavorable conditions for modeling. We identify two separate constraints that limit sampling cadences in ground-based surveys (Moreno et al., 2019): seasonal gaps and limited irregular epochs (sparse irregular observations). For example, object Fairall 9 has 246 observations that cover 1318 days, with two big 100-day gaps. This example demonstrates irregular and unevenly clustered data. The observational time baseline for each object is about 2000 days which is significant for testing our tool on light curves with a substantial length.

[1]For details about the sources see also https://swift.gsfc.nasa.gov/results/bs9mon/

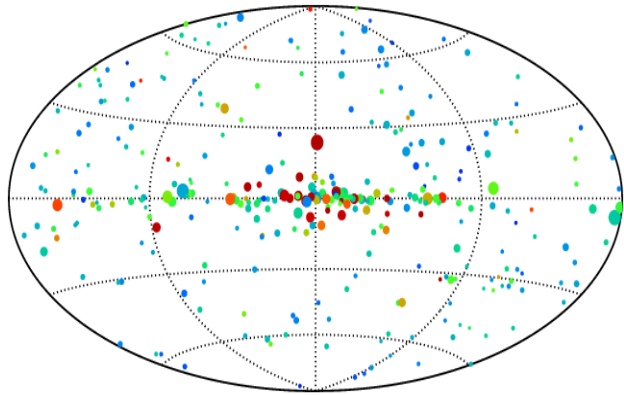

**FIGURE 1** Sky distribution of the 153 AGNs from the 9-Month BAT Survey (Tueller et al., 2008) used for training and testing CNP. Red dots denote soft X-ray sources, blue dots indicate hard X-ray sources, green dots represent stars and the dot diameter is proportional to the source flux (image source: https://swift.gsfc.nasa.gov/results/bs9mon/).

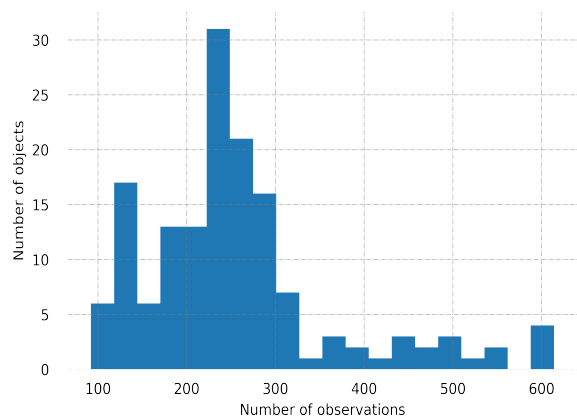

**FIGURE 2** Distribution of the number of observations for the set of 153 light curves in our AGN sample.

## 3 | METHOD

Neural networks have shown good results in modeling stochastic data, without any additional parameters included in processing. Building on previous tools based on Gaussian Processes for AGN light curve analysis and modeling (Kovačević et al., 2019;Shapovalova et al., 2017; Shapovalova et al., 2019), we have implemented the non-parametric CNP method (Garnelo et al., 2018), based purely on data (observations). The CNP has been implemented as supervised training in attempt to approximate unknown functions given only a finite set of observations. This is the first attempt to examine applicability of the CNP for AGN light curve modeling.

Our neural network architecture is developed in `Python` and uses `TensorFlow` (Abadi et al., 2016). It consists of two multi-layer perceptron neural networks and an aggregator. Input data are time in modified Julian date $x_i$ and measured flux in mJy $y_i$.

Our task is to try to find the values of flux in the moments of time where we have gaps in our observations in order to obtain realistic and continuous representation of our light curve. First, we take from our data-set $O$, which consists of $n$ points, a sub-set $O_N$ containing context points, with $N$ indicating the number of points and another set with $m$ unlabeled target points $T$:

$$O = \{(x_i, y_i)\}_{i=0}^{n-1} \subset X \times Y$$
$$O_N = \{(x_i, y_i)\}_{i=0}^{N} \subset O$$
$$T = \{(x_i)\}_{i=n}^{n+m+1} \subset X.$$

Context points are used for training the neural network, whereas the target points are used for making predictions, i.e. modeling the light curve. The context set is passed through an encoder, as illustrated in the algorithm in Figure 3.

The encoder is a multi-layer perceptron (MLP) neural network which produces as output lower dimensional representations (represented with $h$ in Figure 3) $r_i = h((x, y)_i)$ for each context point. The equation notation means that $r_i$ is the value of one function h in different data points $(x, y)_i$. We assume that the outputs are a realization of a stochastic process. The aggregator combines all these representations into one unique representation. In our implementation the aggregator (represented with $a$ in Figure 3) calculates the mean value of representations.

This unique $r$ is meant to have all of the information of the unknown function that we have obtained with these few observations. We then combine this unique representation with our target points $x_t$ (the unlabeled points) and pass it to a decoder. The decoder, which is also a multi-layer perceptron neural network, calculates predictions for each target value and outputs mean and variance of the predicted distribution.

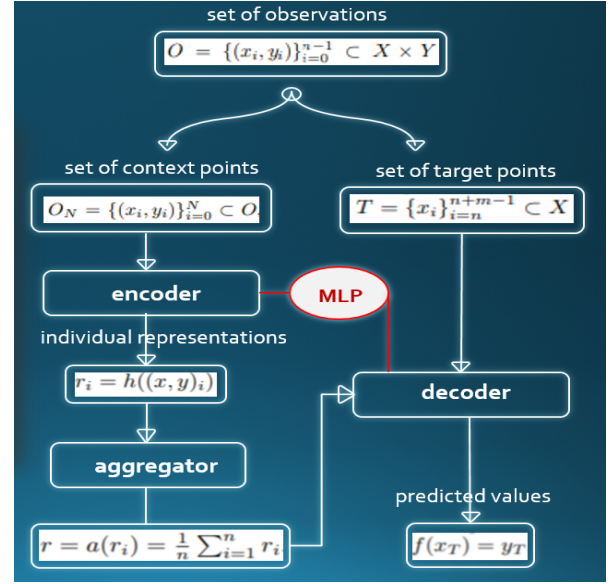

**FIGURE 3** Schematic representation of the Conditional Neural Process, representing data flow. Context points are processed by encoder and aggregator. The output of this process is sent to the decoder along with target points to calculate predictions.

With the mean and variance our, method calculates the log probability of predicted $y$ values, which is used to back-propagate or update parameters (weight and biases) of our model in the next iteration step. The loss function corresponds to the negative log probability of the ground truth targets under the predicted distribution and can be interpreted as the degree to which an event supports a statistical model. [2]

In the first phase of our research, we have used target points for which we know the measured values of flux (to determine how accurate the predictions are by comparing them with real values). In the second phase, we have used equidistant target points to make predictions, which allowed us to generate a more suitable model for further analysis. The aim of our training is to minimize the loss function (Garnelo et al., 2018):

$$\mathcal{L}(\theta) = -\mathbb{E}_{f \sim P}\left[\mathbb{E}_N\left[\log Q_\theta(\{y_i\}_{i=0}^{n-1} | O_N, \{x_i\}_{i=0}^{n-1})\right]\right].$$

As the loss function decreases, the log-likelihood of the target points becomes higher, i.e. predicted distribution becomes a better fit of a statistical model to our data sample. In practice, we use Adam optimizer (Kingma et al., 2017), an algorithm for first-order gradient-based optimization of stochastic objective functions, since it is computationally efficient, has little memory requirements, and is well suited for problems that are large in terms of data. The method is also appropriate for non-stationary objectives and problems with very noisy and sparse gradients (Kingma et al., 2017). This approach allows method to rely only on the empirical data rather then imposing assumption of an analytic prior (Garnelo et al., 2018).

---

[2] A simplified explanation is that the error between the prediction output and the specified target value is measured using loss functions. A loss function indicates how close the algorithm model is to achieving the desired result. The term loss refers to the penalty that the model receives when it fails to produce the expected results. If the deviation of the model is high the loss value will be also very high and vice versa.

CNP method can make good predictions with handful of data. We demonstrated it on artificial sparse sample generated by GP (see Figure 4). The ground truth GP curve is shown as a black dotted line and the context points from this curve that are fed into the model as black dots. The model's predicted mean and variance are shown as solid green line and light green shaded area, respectively, after 100 000 iterations. The next steps for future research of this functionality on real AGN light curve data are described in Section 5.

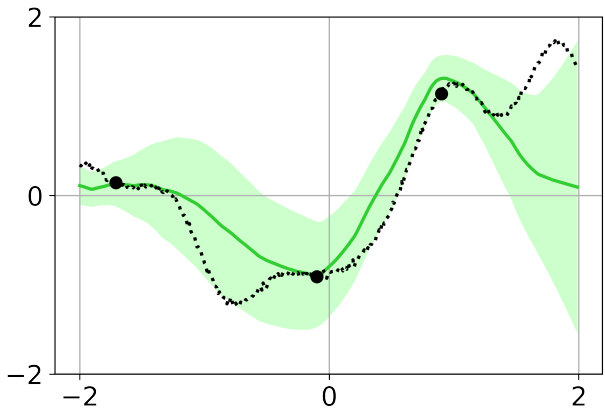

**FIGURE 4** CNP modeling of GP generated data using 100 000 iterations. The light green area is the one sigma confidence interval. Black dots are context points. Black dotted line is the ground truth curve.

Another goal of our work is to make our code adaptable for processing vast amount of data (AGNs and light curves of other sources), expected to be harnessed by LSST surveys. For this purposes we made initial parallelization attempts of our code and tests on high performance computing facility. The simulations were run on the SUPERAST, an HPE ProLiant DL380 Gen10 2x Xeon 4210-S 64GB SFF server located at the Department of Astronomy, Faculty of Mathematics, University of Belgrade (described by Kovačević et al., 2021). There are two computational nodes in SUPERAST. Each node has 40 cores, 128GB RAM, 2TB memory on SSD and 3 DP GFLOPS (DoublePrecision Giga Floating Point Operations Per Second).

The first results shown in Section 5, set the path for future work.

## 4 | RESULTS

Here we give several representative results of AGN light curve modeling after 300 000 iterations that show how the CNP performs with various features of input data sets (Figures 5-9). Some light curves are very difficult to model and each curve has a different structure. In Figures 4-8, dark blue dots denote actual observed values of flux (our context points), whereas light green dots represents predicted values, at target points. The light green shaded band represents the confidence interval, which shrinks after each iterations step. We now further discuss each individual case. Their basic CNP parameters are listed in Table 1.

The first two cases, which are presented in Figures 4 and 5 are dominated by significant gradients in light curves. Figure 5 shows the modeled light curve for 3C 382, a nearby ($z = 0.058$) broad-line radio galaxy (BLRG) hosting a supermassive black hole of $1.0 \pm 0.3 \times 10^9 M_\odot$ (obtained from reverberation mapping Fausnaugh et al., 2017). The observation baseline is 1253 days (Table 1). We can see that our tool was able to model the prevailing gradient changing from ascending to descending, with a prominent peak at 57250 MJD.

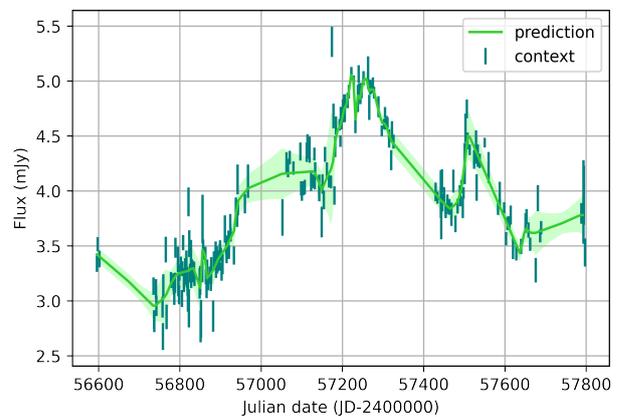

**FIGURE 5** CNP modeling of 3C 382 light curve using 300 000 iterations. The light green area is the one sigma confidence interval. The width of confidence intervals depends on number of context points (see Garnelo et al., 2018, and their Figure 2).The larger number of context points produce narrower confidence bands.

Figure 6 shows application of CNP on the light curve of Fairall 9, a radio–quiet Seyfert type 1 AGN located at $z = 0.0461$ (Lauberts et al., 1989), with a super-massive black hole of $2.55 \pm 0.56 \times 10^8 M_\odot$ (measured from reverberation mapping Peterson et al., 2004). The observed baseline for Fairall 9 is 1318 days. The modeled light curve for this AGN shows extreme gradient changes in a 400-day period. This object has also been processed with observational errors included in pre-processing (the light curve has been normalized with observational errors prior to modelling). The difference for loss function in these two cases was 0.0009 (Table 1). We can see that our model is able to properly learn rapid variations, but after a considerable number of iterations.

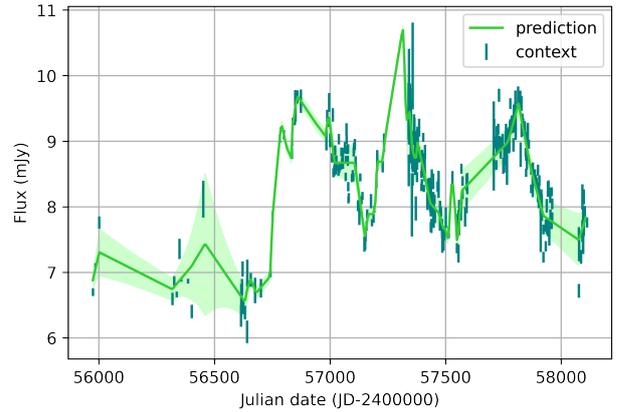

**FIGURE 7** The same as in Figure 4, but for 2MASX J11454045-1827149.

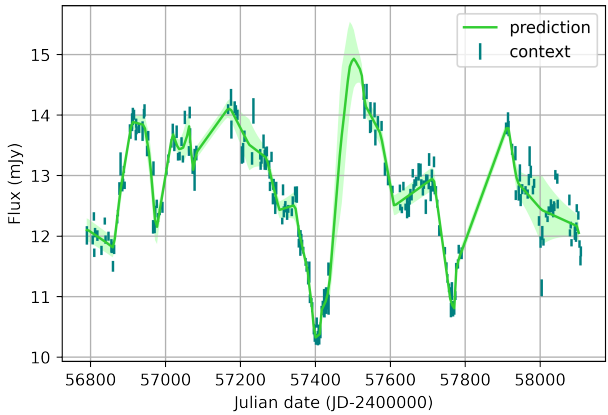

**FIGURE 6** The same as in Figure 4, but for Fairall 9.

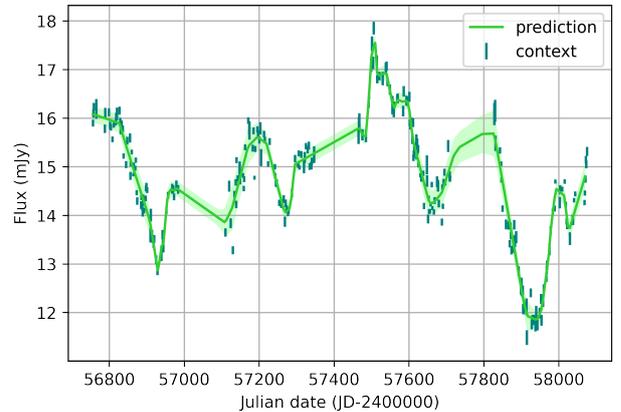

**FIGURE 8** The same as in Figure 4, but for Mrk 509.

Following examples show CNP modeling of light curves with large observational gaps (Figures 7 and 8). The light curve of Seyfert 1 galaxy 2MASX J11454045-1827149 (Figure 7) is an example of a very complex structure with low density of observations in the first 1000 days as opposed to the last 1000 days which are very tightly sampled, with visible flares and successive peaks suggesting possible quasi periodic oscillations. The observed length of 2MASX J11454045-1827149 is 2138 days.

The light curve of the nearby ($z = 0.034$) Seyfert 1 galaxy Markarian 509 (Figure 8), shows well modeled sequential gradient changes (Marshall et al., 2009), despite the irregular density of observations and several gaps. The observed length of the Markarian 509 light curve is 1319 days.

In the final example, the Seyfert 1 galaxy NGC 7469 (Figure 9), we can see a light curve with a convex form which also has quite irregular structure with large gaps, flares, and possible quasi-periodic oscillations. The observed length of the NGC 7469 light curve is 2136 days.

## 5 | DISCUSSION

We faced a variety of challenging issues during testing process. Some of them required dividing data into subsets to overcome big gaps (2MASX J11454045-1827149, Fairall 9, 3C 382, NGC 7469), to eliminate flares (2MASX J11454045-1827149, 3C 382, Fairall 9) or to exclude additional iterations (3C 382, NGC 7469), which increased execution time significantly.

As mentioned in Section 3, loss is the negative log-likelihood and as such it represents a measure of how well the model is able to predict the actual value. The goal is to minimize the error of predictions, hence the loss function decreases with each iteration step. The loss function depends highly on the number of iterations (Table 1). From the examples provided here we can see that the loss function has values between 0.0016 and 0.1775 which suggests that the CNP makes accurate predictions, since loss function is calculated

**TABLE 1** Basic parameters for CNP modeling performance for a selected sub-sample of representative cases. The columns are: object name, number of observations in the light curve (*n*), observation time-baseline (L), CNP algorithm execution time (T) for 300 000 iterations, and the best values for loss function obtained during 300 000 iterations for objects: 3C 382, Fairall 9, Mrk 509, NGC 7469 and 2MASX J11454045-1827149. All objects have been processed without observational errors, with the exception of Fairall 9, where given values present results with and without measurement errors included in processing.

| object | *n* | L (days) | $T(sec)$ | loss |
|---|---|---|---|---|
| 3C 382 | 225 | 1253 | 1735 | 0.004859 |
| Fairall 9 | 246 | 1318 | 1898 | 0.007852 |
| Fairall 9* | 246 | 1318 | 1898 | 0.008732 |
| Mrk 509 | 259 | 1319 | 1857 | 0.001648 |
| NGC 7469 | 264 | 2136 | 1809 | 0.008796 |
| 2MASX J11454045-1827149 | 376 | 2138 | 1917 | 0.177469 |

*Object's light curve modeled with measurement errors included in processing.

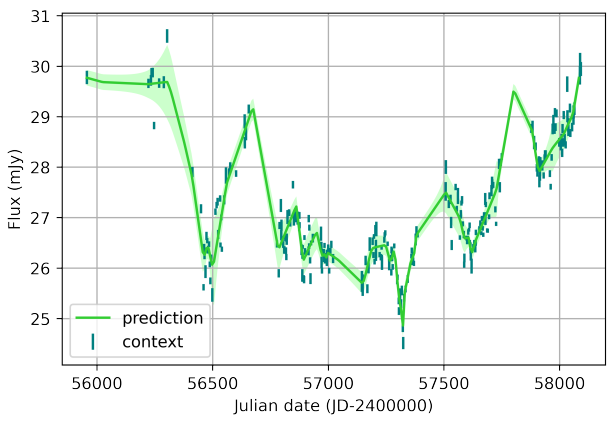

**FIGURE 9** The same as in Figure 4, but for NGC 7469.

as the standard deviation of the multivariate Gaussian using softplus function:

$$\ln(1 + e^x) \in (0, \infty).$$

The CNP scales as $O(n + m)$ at execution time, where *n*, *m* are the number of observations and targets, respectively (as defined in Section 3). GPs, due to difficulty of designing appropriate priors for complex tasks, scale poorly with *n* and *m*, $O((n + m)^3)$ (Garnelo et al., 2018). Therefore, we have tested our code with different setups and neural network configurations, namely changing the number of iterations, layers or target points.

After examining, the results for all 153 processed objects, we observed linear dependency between the loss function and the number of observations (Figure 10). This led us to conclusion that parallelisation of our data could improve CNP efficiency, not only by speedup of the execution, but also by making more accurate predictions. Dividing each light curve in sub-intervals (along the dimension of MJD) would result in less complex structure in each subset, and hence less challenging data for modeling. We have used this approach in CNP modeling for preparing AGN light curves for photometric reverberation mapping (Jankov et al., 2021).

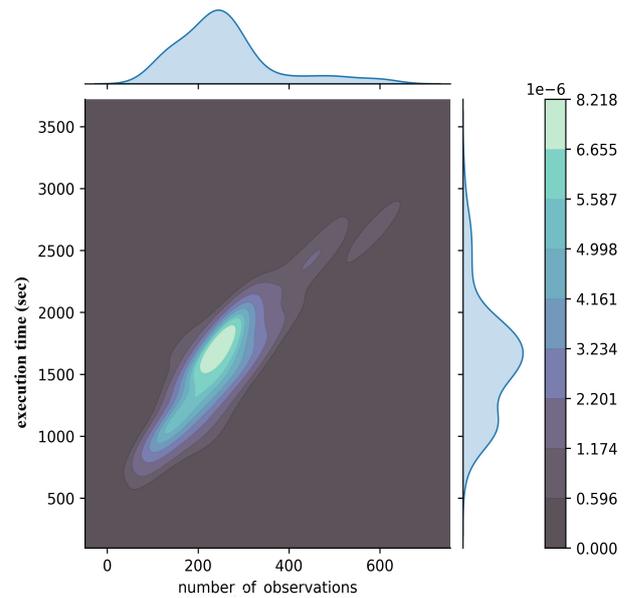

**FIGURE 10** Time of execution in seconds as a function of the number of observations for 300 000 iterations for all 153 processed objects. Marginal distribution of time and number of observations is also shown on the sides of plot. Color bar represents probability density function.

Another goal of our project was code optimisation. We have implemented data parallelism, using module mpi4py (Dalcin et al., 2018) and tested it on a computer cluster. These first tests have been made on a small scale (2, 4 and 8 cores) to determine possible issues and resolve them prior to more comprehensive examination.

The best result for speedup (63%) has been obtained while testing on 4 cores for 300 iterations (Figure 11). Another interesting manifestation that emerged during testing with 3000 iterations is that the execution time decreased up to 4 cores, and with this configuration reached its minimum. Further decrees in execution time did not appear, even with usage of additional processing units (Figure 12).

Our initial hypothesis was that this phenomena could be a representation of Amdahl's law (Amdahl, 1967), which refers to the speedup gained as a result of the parallelization (Karbowski, 2008) i.e.,

$$S(s, p) = \frac{1}{\beta(s,1)+[1-\beta(s,1)]/p} \quad (1)$$

where $S(s, p)$ is application speedup (scalability) which is dependent on the problem size $s$ and the number of processors $p$. The sequential fraction of the code is denoted with $\beta(s, 1)$, while $T(s, p)$ is the total real-time of the execution of the program on a machine with $p$ processors, and $T(s, 1)$ represents the execution of the program on one processor. Amdahl's law limits application speedup for large p.

$$S(s, p) \xrightarrow{p \to \infty} \frac{1}{\beta(s, 1)} \quad (2)$$

In attempt to investigate this possibility, we have calculated Amdahl's law assuming our testing could be caused by it. If we take from the results represented in Figure 11 that speedup on 4 cores for 3000 iterations is equal to 2.104 (taking that the execution time on one core $T(s, 1)$ is 202 seconds, divided by 96 seconds which is the execution time on four cores $T(s, p)$ for the same number of iterations and equivalent input data), then we can calculate from equation (1) that the serial (sequential) part of our application is $\beta(s, 1) = 0.3$ and the parallel part is 0.7. We conclude that 70% of our current code could be parallelized.

This should be taken with caution, since there could be other factors that influence execution time (server setup, module configuration, etc). More conclusive results require further development, additional testing, and analysis.

One important matter which should be addressed in the future research, aside from parallelization is estimating the best number of iterations for obtaining the minimal value for the loss function. Figure 13 shows how the loss function changes during modeling light curves with 300 000 iterations. It appears that the loss function sometimes reaches its minimum very fast (after several thousand iterations), suggesting that further training process is unnecessary and time of execution for this cases could be significantly shorter. In addition, it would be interesting to examine the optimal number of batches for gaining desirable modeling accuracy, since that could be the means for shortening execution time even further.

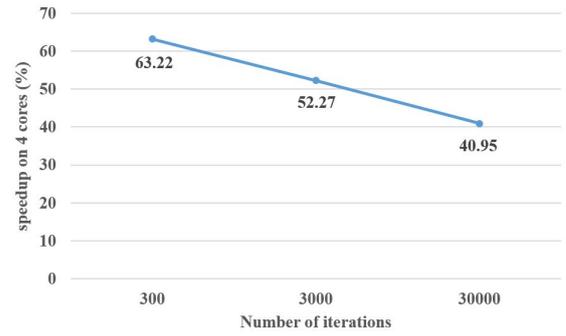

**FIGURE 11** Speedup on four cores for different number of iterations. Speedup decreases with the increase of the number of iterations.

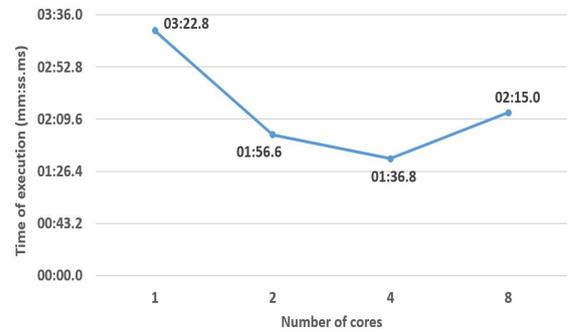

**FIGURE 12** Speedup for 3000 iterations for 1, 2, 4, and 8 cores, tested on a computer cluster. There is a visible increase in the time of execution after the number of cores reaches 4, which requires further analysis. Time of execution is in format (mm:ss.ms).

Our future plans also include testing of the CNP prediction of light curve behavior in large gaps by removing some of the data from the long, well-sampled light curves. We plan to use as a testbed well sampled light curves of NGC 5548, NGC 4151, NGC 4593, and Mrk 509 discussed by Edelson et al. (2019). The initial testing for this functionality have been made with artificially generated data described in Section 3.

## 6 | CONCLUSION

In this work, we have implemented the Conditional Neural Process and examined its applicability to modeling AGN light curves. We tested it on a sample of 153 AGNs from ASAS-SN, with some time baselines exceeding 2000 days. The first results are encouraging and show it to be a good alternative to the previous methods. CNP makes accurate predictions after training on various sets of real AGN light curve data, without any additional parameters or simulated noise. Time of execution grows with a number of context points (observations). Further optimisation and parallelization of the code are necessary for more efficient processing of a substantial amount of data.

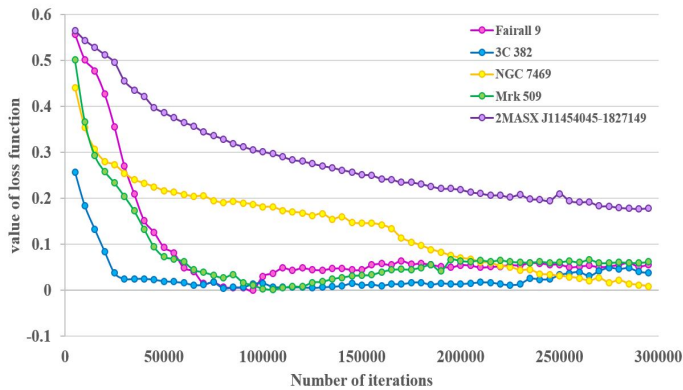

**FIGURE 13** Comparison of CNP loss function values for sample of objects, with respect to the number of CNP iterations. The legend shows color coding of loss functions according to the object.

The results suggest that CNP is an adequate tool for modeling stochastic data with random gradient change, but at a cost of a larger number of iterations and hence significant execution time. CNP represents a good alternative to Gaussian Process modeling, which has been used for modeling of AGN light curves in order to capture the variation of local patterns of their time evolving delay (Kovačević et al., 2018). Even though CNP modeling has less then optimal execution time, it is comparable with other machine learning algorithms based on GP algorithms, since they require determining the most adequate kernel, which could also be computationally expensive. Furthermore, since CNP is non-parametric and does not require preprocessing of input data, it is applicable to other astronomical objects. The ability of CNP to scale to complex functions and large data sets indicates encouraging possibilities for future use in AGN time-domain analysis.

Considering that the CNP could be more efficient in modeling our data, the results indicate that it would be advisable to make other optimisations of our code in future work, so that it could be useful for the long-time baseline surveys such as the LSST (Ivezić et al., 2019) which will harvest millions of AGN light curves. Our plans include further work on parallelization, comparison with other methods for modeling AGN light curves, and implementation of mass processing.

## ACKNOWLEDGMENTS

The authors express their gratitude to the Reviewer for insightful remarks and corrections, which aided in the presentation of the paper. This project is graciously supported by a grant from the 2021 LSST Corporation Enabling Science Call for Proposals, and funding provided by Faculty of Mathematics University of Belgrade (the contract 451-03-9/2021-14/200104) and Astronomical Observatory (the contract 451-03-68/2020-14/200002), through the grants by the Ministry of Education, Science, and Technological Development of the Republic of Serbia. A. K. and L. Č. P. acknowledge the support by Chinese Academy of Sciences President's International Fellowship Initiative (PIFI) for visiting scientist. D.I. acknowledges the support of the Alexander von Humboldt Foundation. This research was possible through the use of data available in the All-Sky Automated Survey for Supernovae (ASAS-SN) database.